\documentclass[copyright]{eptcs}
\providecommand{\event}{T. Neary, D. Woods, A.K. Seda and N. Murphy (Eds.):\\ The Complexity of Simple Programs 2008.}
\providecommand{\volume}{1}
\providecommand{\anno}{2009}
\providecommand{\firstpage}{185}
\usepackage{breakurl}

\usepackage{bm}
\usepackage{latexsym}
\usepackage{amssymb}
\usepackage{amsmath}
\usepackage{amsthm}
\usepackage{verbatim}
\usepackage{graphicx}

\newtheorem{theorem}{Theorem}
\newtheorem{lemma}{Lemma}
\newtheorem{example}{Example}
\newtheorem{corollary}{Corollary}

\newtheorem{definition}{Definition}

\newcommand{\makeset}[2]{\ensuremath{ \{ #1 \: | \: #2 \} }}

\newcommand{\family}[1]{\textsf{\small \textup{\textit{#1}}}}

\renewcommand {\epsilon}{\varepsilon}

\newcommand{\qunknown}{{?}}
\newcommand{\qv}[1]{#1}
\newcommand{\qvleft}[1]{#1^{\nwarrow}}
\newcommand{\qvright}[1]{#1^{\nearrow}}
\newcommand{\qvmem}[2]{#1^{#2}}

\title{Representing a {P}-complete problem by small trellis automata}%
\author{Alexander Okhotin
\institute{Department of Mathematics, University of Turku, Turku FIN--20014, Finland, \emph{and} Academy of Finland.\thanks{Supported by the Academy of Finland under grant 118540.}}
\email{alexander.okhotin@utu.fi}
}
\def\titlerunning{Representing a {P}-complete problem by small trellis automata}
\def\authorrunning{A. Okhotin}

\begin{document}
\maketitle
\sloppy

%\begin{frontmatter}

\begin{abstract}
A restricted case of the Circuit Value Problem
known as the Sequential NOR Circuit Value Problem
was recently used to obtain
very succinct examples
of conjunctive grammars, Boolean grammars and language equations
representing P-complete languages
(Okhotin,
\href{http://dx.doi.org/10.1007/978-3-540-74593-8_23}
{``A simple P-complete problem and its representations by language equations''},
MCU 2007).
In this paper, a new encoding of the same problem is proposed,
and a trellis automaton (one-way real-time cellular automaton)
with 11 states
solving this problem is constructed.
\end{abstract}
%\end{frontmatter}

\section{Introduction}

Many kinds of automata and formal grammars
have the property that
all sets they define are contained
in some complexity class $\mathcal{C}$,
and at the same time they can define
some particular set complete for $\mathcal{C}$
(in the sense that every set in $\mathcal{C}$
can be reduced to that set).
When $\mathcal{C}$ is the family of recursively enumerable sets
and many-one reductions are considered,
such models are known as \emph{computationally universal},
and the same phenomenon occurs in formalisms
of widely different expressive power.

For instance, for linear context-free grammars
it is known that all languages they generate
are contained in \family{NLOGSPACE},
and Sudborough~\cite{Sudborough}
constructed a small example of a linear context-free grammar
that generates an \family{NLOGSPACE}-complete language.
Such a result is essential, in particular,
to understand the complexity of parsing these grammars.
Having a succinct example is especially good,
as it shows the refined essense of the expressive power
of linear context-free grammars
in an easily perceivable form.

\begin{comment}
Such an example witnesses
understanding of the expressive power of the formalism,
and communicates refined knowledge on its power.
Accordingly, the grammar of Sudborough~\cite{Sudborough}
witnesses that linear context-free grammars are understood,
\end{comment}

Thus for every such formalism
(as long as the formalism is of any importance),
it is interesting to obtain
a succinct representation of a complete problem.
Results of this kind
have recently been obtained by the author \cite{BooleanPComplete}
with respect to another two families of formal grammars:
\emph{conjunctive grammars} \cite{Conjunctive}
and \emph{Boolean grammars} \cite{BooleanGrammars},
which are extensions of the context-free grammars
with Boolean operations.
The languages generated by these grammars
are contained in $\family{DTIME}(n^3) \subset \family{P}$,
and grammars generating $\family{P}$-complete languages
with 8 and 5 rules, respectively,
were constructed~\cite{BooleanPComplete}.
The underlying idea of the construction
was a specific new variant of the Circuit Value Problem,
which maintains \family{P}-completeness
and is particularly suitable for representation
by these grammars.

This paper is concerned with finding succinct representations
of \family{P}-complete languages
for another important model: the \emph{trellis automata}.
Trellis automata are one of the simplest,
perhaps \emph{the} simplest kind of cellular automata,
and are known as \emph{one-way real-time cellular automata}
in the standard nomenclature.
The first results on their expressive power are due to
Smith~\cite{Smith}, Dyer~\cite{Dyer}
and Culik et al.~\cite{CulikGruskaSalomaa}.
As a trellis automaton uses space $n$
and makes $\Theta(n^2)$ transitions,
every language it recognizes is in $\family{P}$;
the existence of a trellis automaton
accepting a \family{P}-complete language
was demonstrated by Ibarra and Kim \cite{IbarraKim},
though no explicit construction was presented.
A linear conjunctive grammar
generating an encoding of the Circuit Value Problem
for a \family{P}-complete problem was constructed
by the author \cite{LinearHardest},
and as a part of this proof,
a construction of a 45-state trellis automaton
over a 9-letter alphabet was given.

This paper aims to construct a new trellis automaton
solving a different \family{P}-complete problem,
this time with the goal of minimizing the number of states.
The problem is the same variant of the Circuit Value Problem
as in the previous paper~\cite{BooleanPComplete},
though this time a new encoding is defined.
With the proposed encoding, the problem may be solved
by an 11-state trellis automaton
over a 2-letter alphabet.
A full construction will be given and explained.

\section{Trellis automata and conjunctive grammars}

Trellis automata can be equally defined
by their cellular automata semantics
(using evolution of configurations)
and through the trellis representing their computation.
According to the latter approach,
due to Culik et al.~\cite{CulikGruskaSalomaa},
a trellis automaton
processes an input string of length $n \geqslant 1$
using a uniform triangular array of $\frac{n(n+1)}{2}$ processor nodes,
as presented in the figure below. %Figure~\ref{f:trellis_automata}.
Each node computes a value from a fixed finite set $Q$.
The nodes in the bottom row obtain their values
directly from the input symbols
using a function $I : \Sigma \to Q$.
The rest of the nodes
compute the function $\delta : Q \times Q \to Q$
of the values in their predecessors.
The string is accepted
	if and only if
the value computed by the topmost node
belongs to the set of accepting states $F \subseteq Q$.
This is formalized in the following definition.

\newlength{\taleft}
\setlength{\taleft}{\textwidth}
\addtolength{\taleft}{-4.7cm}
\begin{definition}
A trellis automaton is a quintuple
$M = (\Sigma, Q, I, \delta, F)$, in which: \\[0.8mm]
\begin{minipage}{\taleft}%{8.1cm}
\begin{itemize}
\item	$\Sigma$ is the input alphabet,
\item	$Q$ is a finite non-empty set of states, %(of processing units),
\item	$I : \Sigma \to Q$ is a function that sets the initial states,
\item	$\delta : Q \times Q \to Q$ is the transition function,
	and
\item	$F \subseteq Q$ is the set of final states.
\end{itemize}
\end{minipage}~\hspace*{2mm}~\begin{minipage}{5.3cm}
\includegraphics[scale=1.2]{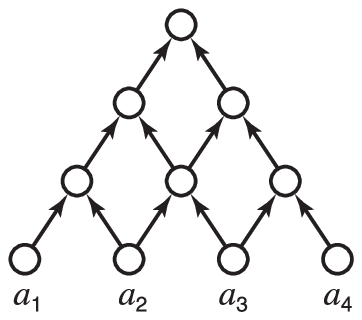}
\end{minipage} \\[0.8mm]
The result of the computation on a string $w \in \Sigma^+$
is denoted by $\Delta : \Sigma^+ \to Q$,
which is defined inductively as $\Delta(a)=I(a)$ and
$\Delta(a w b)=\delta(\Delta(aw), \Delta(wb))$,
for any $a, b \in \Sigma$ and $w \in \Sigma^*$.
Then the language recognized by the automaton
is $L(M)=\makeset{w}{\Delta(w) \in F}$.
\end{definition}

Trellis automata are known to be equivalent
to \emph{linear conjunctive grammars}~\cite{LinearAutomata}.
These grammars are subclass of \emph{Boolean grammars},
which are a generalization of the context-free grammars
with explicit Boolean operations.
In addition to the implicit disjunction
represented by multiple rules for a single nonterminal,
which is the only logical operation expressible in context-free grammars,
Boolean grammars allow both conjunction and negation in the formalism of rules.

\begin{definition}
A Boolean grammar \cite{BooleanGrammars}
is a quadruple $G=(\Sigma, N, P, S)$,
in which
\begin{itemize}
\item
	$\Sigma$ and $N$ are disjoint finite nonempty sets
	of terminal and nonterminal symbols, respectively;
\item
	$P$ is a finite set of rules of the form
	\begin{equation} \label{rule_in_Boolean_grammar}
		A \to \alpha_1 \& \ldots \& \alpha_m \& \lnot \beta_1 \& \ldots \& \lnot \beta_n
		\quad (A \in N, \; m+n \geqslant 1, \; \alpha_i, \beta_j \in (\Sigma \cup N)^*),
	\end{equation}
\item	$S \in N$ is the start symbol of the grammar.
\end{itemize}
For each rule (\ref{rule_in_Boolean_grammar}), the objects
$A \to \alpha_i$ and $A \to \lnot \beta_j$ (for all $i, j$) are called conjuncts,
positive and negative respectively.
\end{definition}

Intuitively, a rule (\ref{rule_in_Boolean_grammar}) can be read as
\emph{``if a string satisfies the syntactical conditions $\alpha_1, \ldots, \alpha_m$ and does not satisfy
any of the syntactical conditions $\beta_1, \ldots, \beta_n$, then this string satisfies the
condition represented by the nonterminal $A$''.}
This intuitive interpretation is formalized
by the following system of \emph{language equations},
in which the nonterminal symbols
represent the unknown languages,
and for every $A \in N$, there is an equation
\begin{equation*}
	A
		=
	\bigcup_{A \to \alpha_1 \& \ldots \& \alpha_m \& \lnot \beta_1 \& \ldots \& \lnot \beta_n \in P}
	\bigg[
		\bigcap_{i=1}^m \alpha_i
		\;\cap\;
		\bigcap_{j=1}^n \overline{\beta_j}
	\bigg].
\end{equation*}
Then the languages generated by the nonterminals of the grammar
are defined by the corresponding components
of a certain solution of this system.
In the simplest definition, the system must have a unique solution,
with some further restriction \cite{BooleanGrammars}.
According to this definition, some grammars,
such as $S \to S$ and $S \to \lnot S$,
are deemed invalid, but in practice
every reasonably written grammar satisfies the definition.
Consider the following example:

\begin{example}[\cite{NineOpenProblems}] \label{abc_neg_example}
The following Boolean grammar generates the language
\makeset{a^m b^n c^n}{m,n \geqslant 0, m \neq n}:
\begin{equation*}\begin{array}{rcl}
	S &\to& AB \& \lnot DC \\
	A &\to& aA \ | \ \epsilon \\
	B &\to& bBc \ | \ \epsilon \\
	C &\to& cC \ | \ \epsilon \\
	D &\to& aDb \ | \ \epsilon
\end{array}\end{equation*}
\end{example}
The rules for the nonterminals $A$, $B$, $C$ and $D$ are context-free,
and so, according to the intuitive semantics,
they should generate the languages
$L(A)=a^*$,
$L(B)=\makeset{b^n c^n}{n \geqslant 0}$,
$L(C)=c^*$ and
$L(D)=\makeset{a^m b^m}{m \geqslant 0}$.
Then the propositional connectives in the rule for $S$
specify the following combination
of the conditions given by $AB$ and $DC$:
\begin{equation*}
	\underbrace{\makeset{a^n b^m c^m}{m,n \geqslant 0, m \neq n}}_{L(S)}
	=
\end{equation*}
\begin{equation*}
	\makeset{a^i b^j c^k}{j=k \text{\ and\ }i \neq j}
	=
%	L(AB) \cap \overline{L(DC)}
	\underbrace{\makeset{a^i b^j c^k}{j=k}}_{L(AB)}
	\cap
	\overline{\underbrace{\makeset{a^i b^j c^k}{i=j}}_{L(DC)}}.
\end{equation*}

A Boolean grammar is called a \emph{conjunctive grammar}
if the negation is never used, that is,
$n=0$ for every rule (\ref{rule_in_Boolean_grammar}).
A conjunctive grammar is a \emph{context-free grammar}
if neither negation nor conjunction are allowed,
that is, $m=1$ and $n=0$ for all rules.
Similarly to the context-free case,
a Boolean (conjunctive) grammar
is called \emph{linear Boolean} (\emph{linear conjunctive})
if the body of every conjunct may contain
at most one reference to a nonterminal symbol,
that is, $\alpha_i,\beta_j \in \Sigma^* \cup \Sigma^* N \Sigma^*$
for each rule (\ref{rule_in_Boolean_grammar}).

\begin{example}[\cite{Conjunctive}]
The following linear conjunctive grammar
generates the language \makeset{wcw}{w \in \{a, b\}^*}:
\begin{equation*}\begin{array}{rcl}
	S &\to& C \& D \\
	C &\to& aCa \ | \ aCb \ | \ bCa \ | \ bCb \ | \ c \\
	D &\to& aA \& aD \ | \ bB \& bD \ | \ cE \\
	A &\to& aAa \ | \ aAb \ | \ bAa \ | \ bAb \ | \ cEa \\
	B &\to& aBa \ | \ aBb \ | \ bBa \ | \ bBb \ | \ cEb \\
	E &\to& aE \ | \ bE \ | \ \epsilon
\end{array}\end{equation*}
\end{example}

It is known that linear conjunctive grammars
and linear Boolean grammars
generate the same family of languages.
Furthermore, as already announced above,
they are computationally equivalent to trellis automata:

\begin{theorem}[Okhotin~\cite{LinearAutomata}] \label{LinM_and_TA_equivalence_theorem}
A language $L \subseteq \Sigma^+$ is generated by a linear conjunctive grammar
	if and only if
$L$ is recognized by a trellis automaton.
These representations can be effectively transformed into each other.
\end{theorem}

In particular,
the conversion of a trellis automaton to a linear conjunctive grammar
can be done quite straightforwardly
by taking a nonterminal $A_q$ for each state $q$ of the automaton
and adding the rules
\begin{equation*}
	A_q \to b A_{q''} \& A_{q'} c
		\quad (\text{for all $q', q'' \in Q$ with $q=\delta(q',q'')$
		and for all $b,c \in \Sigma$}),
\end{equation*}
as well as a rule $A_{I(a)} \to a$ for every $a \in \Sigma$.
If there is a unique accepting state $q$,
then $A_q$ may be taken for a start symbol,
and otherwise a new start symbol has to be defined.

In this way an automaton with $n$ states and $m$ letters
is converted to a grammar with at most $n+1$ nonterminal symbols
and at most $m^2n^2+m+n$ rules.
A more complicated conversion is known~\cite{LinearNonterminals},
which always produces a grammar with 2 nonterminals.
However, the number of rules in the grammar becomes enormous,
so this result will not produce any succinct representations.

\section{Sequential NOR Circuit Value Problem}

A circuit is an acyclic directed graph, in which
the incoming arcs in every vertex are considered ordered,
every source vertex is labelled with a variable
from a certain set $\{x_1, \ldots, x_m\}$ with $m \geqslant 1$,
each of the rest of the vertices
is labelled with a Boolean function of $k$ variables
	(where $k$ is its in-degree),
and there is a unique sink vertex.
For every Boolean vector of input values $(\sigma_1, \ldots, \sigma_m)$
assigned to the variables, the value computed at each gate is defined as the value
of the function assigned to this gate on the values computed in the predecessor gates.
The value computed at the sink vertex is the
output value of the circuit on the given input.

The \emph{Circuit Value Problem} (CVP) is stated as follows:
given a circuit with gates of two types, $f_1(x)=\lnot x$ and $f_2(x, y)=x \land y$,
and given a vector $(\sigma_1, \ldots, \sigma_m)$
of input values assigned to the variables ($\sigma_i \in \{0, 1\}$),
determine whether the circuit evaluates to 1 on this vector.
The pair \emph{(circuit, vector of input values)} is called an instance of CVP.
This is the fundamental problem complete for \family{P}
with respect to logarithmic-space many-one reductions,
which was proved by Ladner \cite{Ladner}.
A variant of this problem is the \emph{Monotone Circuit Value Problem} (MCVP),
in which only conjunction and disjunction gates are allowed.
As shown by Goldschlager \cite{Goldschlager}, MCVP remains \family{P}-complete.

A multitude of other particular cases of CVP are known to be \family{P}-complete
\cite{PCompletenessTheory}.
Let us consider one particular variant of this standard computational problem.
A \emph{sequential NOR circuit} is a circuit satisfying the following conditions:
\begin{itemize}
\item	The notion of an input variable is eliminated, and the circuit is deemed to
	have a single source vertex, which, by definition, assumes value 1.
\item	A single type of gate is used. This gate implements \emph{Peirce's arrow}
	$x \downarrow y=\lnot(x \lor y)$, also known as the NOR function.
	It is well-known that every Boolean function can
	be expressed as a formula over this function only.
\item	The first argument of every $k$-th NOR gate
	has to be its direct predecessor, the $(k-1)$-th gate,
	while the second argument can be any previous gate.
	Because of that, these gates will be called \emph{restricted NOR gates}.
\end{itemize}
The problem of testing whether such a circuit evaluates to 1 is called
the \emph{Sequential NOR Circuit Value Problem},
and it has recently been proved by the author~\cite{BooleanPComplete}
that it remains \family{P}-complete.

\begin{theorem}[\cite{BooleanPComplete}]
Sequential NOR CVP is \family{P}-complete.
\end{theorem}

The idea of the proof
is to simulate unrestricted conjunction and negation gates
by sequences of restricted NOR gates.
An unrestricted negation gate of the form $C_i=\lnot C_j$
can be simulated by two gates:
	$C_i=C_{i-1} \downarrow C_1$
and
	$C_{i+1}=C_i \downarrow C_j$.
The gate $C_1$ is assumed to have value 1,
so $C_i$ will always evaluate to $0$.
Then $C_{i+1}$ computes $\lnot (0 \lor C_j) = \lnot C_j$.

Similarly, a conjunction of $C_j$ and $C_k$
is represented by five restricted NOR gates:
	$C_i=C_{i-1} \downarrow C_1$,
	$C_{i+1}=C_i \downarrow C_j$,
	$C_{i+2}=C_{i+1} \downarrow C_1$,
	$C_{i+3}=C_{i+2} \downarrow C_k$ and
	$C_{i+4}=C_{i+3} \downarrow C_{i+1}$.
Here $C_i$ and $C_{i+2}$ both evaluate to $0$,
$C_{i+1}$ and $C_{i+3}$ compute $\lnot C_j$ and $\lnot C_k$, respectively,
and then the value of $C_{i+4}$ is $C_j \land C_k$.

\section{Representation by language equations} \label{section_language_equations}

The first \family{P}-completeness results
established using Sequential NOR CVP
referred to of language equations of different kinds,
as well as conjunctive and Boolean grammars \cite{BooleanPComplete}.
These results will be briefly explained in this section;
for more explanations the reader is referred to the cited extended abstract.

The expressive means of Boolean grammars
are centered at \emph{recursive definition} of languages,
where the membership of a string in the language
is defined via the membership of shorter strings
in the languages generated by nonterminals of this grammar.
An encoding of the given \family{P}-complete problem
that is particularly suited to recursive definition
can be defined as follows~\cite{BooleanPComplete}.

Every sequential NOR circuit shall be represented
as a string over the alphabet $\{a, b\}^*$.
Consider any such circuit
\begin{align*}
	C_1 & = 1 \\
	C_2 & = C_1 \downarrow C_1 \\
	C_3 & = C_2 \downarrow C_{j_3} \\
	& \vdots \\
	C_{n-1} & = C_{n-2} \downarrow C_{j_{n-1}} \\
	C_n & = C_{n-1} \downarrow C_{j_n}
\end{align*}
where $n \geqslant 1$ and $1 \leqslant j_i < i$ for all $i$.
The gate $C_1$ is represented by the empty string.
Every restricted NOR gate $C_i=C_{i-1} \downarrow C_{j_i}$
is represented as a string $a^{i-j_i-1} b$.
The whole circuit is encoded as a concatenation of these representations
in the reverse order, starting from the circuit $C_n$ and ending with $\ldots C_2 C_1$:
\begin{equation*}
	a^{n-j_n-1} b\,
	a^{(n-1)-j_{n-1}-1} b\,
	\ldots\,
	a^{3-j_3-1} b\,
	a^{2-j_2-1} b
\end{equation*}

The language of correct circuits that have value 1 has the following fairly
succinct definition:
\begin{equation*}\begin{split}
	\{a^{n-j_n-1} b a^{(n-1)-j_{n-1}-1} b \ldots a^{3-j_3-1} b a^{2-j_2-1} b \: | \:
		\text{$n \geqslant 0$ and $\exists y_0, y_1, \ldots, y_n$, s.t.} \\
	\text{$y_1=y_n=1$ and $\forall i$ ($2 \leqslant i \leqslant n$),
	$1 \leqslant j_i < i$ and $y_i=\lnot(y_{i-1} \lor y_{j_i})$} \}
\end{split}\end{equation*}
This is a \family{P}-complete language, and it has a simple structure
that resembles the examples common in formal language theory.
As it will now be demonstrated, this set can indeed be very succinctly defined
by language-theoretic methods.

The set of well-formed circuits that have value 1
(that is, the yes-instances of the CVP)
can be defined inductively as follows:
\begin{itemize}
\item	The circuit $\epsilon$ has value 1.
\item	Let $a^m b w$ be a syntactically correct circuit.
	Then $a^m b w$ has value 1
		if and only if
	both of the following statements hold:
	\begin{enumerate}
	\item	$w$ is \emph{not} a circuit that has value 1
		(in other words, $w$ is a circuit that has value 0);
	\item	$w$ is in $(a^* b)^m u$, where $m \geqslant 0$ and
		$u$ is \emph{not} a circuit that has value 1
		(that is, $u$ is a circuit that has value 0).
	\end{enumerate}
\end{itemize}

Checking the representation
	$a^m b (a^* b)^m u$
requires matching the number of $a$s in the beginning of the string to the number
of subsequent blocks $(a^* b)$,
which can naturally be specified
by a context-free grammar for the following language:
\begin{equation} \label{the_language_L0}
	L_0=\bigcup_{m \geqslant 0} a^m b (a^* b)^m
\end{equation}
To be precise, the language $L_0$ is linear context-free
and deterministic context-free;
furthermore, there exists an LL(1) context-free grammar for this language.

Using $L_0$ as a constant,
one can construct the following language equation,
which is the exact formal representation of the above definition
of the set of circuits that have value 1:
\begin{equation}\label{nonmonotone_equation_for_CVP_plus_junk}
	X = \overline{a^* b X} \cap \overline{L_0 X}
\end{equation}
According to the definition, a string that is a well-formed circuit
has value 1 if and only if it satisfies
(\ref{nonmonotone_equation_for_CVP_plus_junk}).

The equation (\ref{nonmonotone_equation_for_CVP_plus_junk})
can be directly transcribed as the following Boolean grammar:
\begin{quote}
	\( S \to \lnot A b S \& \lnot C S \) \\
	\( A \to a A \ | \ \epsilon \) \\
	\( C \to a C A b \ | \ b \)
\end{quote}

Note that this grammar does not require
a string to be a valid description of a circuit.
For strings that are not well-formed circuits,
the equation (\ref{nonmonotone_equation_for_CVP_plus_junk})
naturally specifies \emph{something},
and some of these strings will be in the solution and some will not.
It would not be difficult at all
to specify syntactical correctness of a circuit
within the grammar.
However, that would lead to a larger grammar,
while the given small grammar
is already sufficient for a \family{P}-completeness argument.

\begin{theorem}[\cite{BooleanPComplete}]
There exists a 5-rule Boolean grammar
that generates a \family{P}-complete language.
\end{theorem}

A very similar construction works without negation.
Let $T$ and $F$ be nonterminals representing
circuits that have value 1 and 0, respectively.
Then these languages can be defined recursively
by the following conjunctive grammar:
\begin{quote}
	\( T \to A b F \& C F \ | \ \epsilon \) \\
	\( F \to A b T \ | \ C T \) \\
	\( A \to a A \ | \ \epsilon \) \\
	\( C \to a C A b \ | \ b \)
\end{quote}

\begin{theorem}[\cite{BooleanPComplete}]
There exists an 8-rule conjunctive grammar
that generates a \family{P}-complete language.
\end{theorem}

\section{Another encoding of circuits} \label{section_encoding}

The encoding of sequential NOR circuits
defined in the previous section
was particularly suited for Boolean grammars.
However, it does not go well with trellis automata,
as they cannot represent concatenation of languages~\cite{Terrier}.

Another encoding of circuits will now be defined.
Again, circuits will be represented
by strings over the alphabet $\{a,b\}$.
Consider any sequential NOR circuit
\begin{align*}
	C_1 & = 1 \\
	C_2 & = C_1 \downarrow C_1 \\
	C_3 & = C_2 \downarrow C_{j_3} \\
	& \vdots \\
	C_{n-1} & = C_{n-2} \downarrow C_{j_{n-1}} \\
	C_n & = C_{n-1} \downarrow C_{j_n}
\end{align*}
where $n \geqslant 2$ and $1 \leqslant j_i < i$ for all $i$.
The gates $C_1$ and $C_2$
are represented by strings $a$ and $b$, respectively.
Every restricted NOR gate $C_i=C_{i-1} \downarrow C_{j_i}$ with $i \geqslant 3$
is represented as a string $b a^{j_i}$.
The whole circuit is encoded
as a concatenation of these representations in the reverse order,
starting from the gate $C_n$
and ending with $\ldots C_3 C_2 C_1$.
The encoding continues with a letter $b$
and a suffix $b^n$ representing the work space
needed by the trellis automaton
to store the computed values of the gates:
\begin{equation*}
	\underbrace{ba^{j_n}
	a^{j_{n-1}}
	\ldots
	ba^{j_4}
	ba^{j_3}
	b
	a}_{\text{gate descriptions}}
	b
	\underbrace{b \ldots b}_{\text{$b^n$: work space}}
\end{equation*}
The set of syntactically correct circuit descriptions
can be formally defined as follows:
\begin{equation*}
	L
		=
	\makeset{b a^{j_n} \, ba^{j_{n-1}} \ldots ba^{j_3} \, b \, a \, b \, b^n}%
		{\text{$n \geqslant 2$ and $1 \leqslant j_i < i$ for each $i$}}.
\end{equation*}
The language of correct descriptions of circuits
that evaluate to 1
has the following fairly succinct definition:
\begin{multline*}
	L_1
		=
	\{ba^{j_n} ba^{j_{n-1}} \ldots ba^{j_3} b ab \, b^n \: | \:
		\text{$n \geqslant 2$ and $\exists x_1, x_2, \ldots, x_n$, s.t.} \\
	\text{$x_1=x_n=1$ and for all $i$ ($1 \leqslant i \leqslant n$),
	$1 \leqslant j_i < i$ and $x_i=\lnot(x_{i-1} \lor x_{j_i})$} \}.
\end{multline*}
This is a \family{P}-complete language
and it has a simple structure
that resembles the examples common in formal language theory.
As it will now be demonstrated, this set can indeed be very succinctly defined
by language-theoretic methods.

\section{Construction of a trellis automaton} \label{section_construction}

The goal is to construct a trellis automaton
that accepts a string from $L$
if and only if it is in $L_1$.
Thus the behaviour of the automaton
on strings from $\{a,b\}^+ \setminus L$ is undefined,
and the actual language it recognizes is different from $L_1$.
As in the case of Boolean grammars,
it would not be difficult to check the syntax by the automaton.
However, disregarding the strings not in $L$
results in a simpler construction and in fewer states.
%The trellis automaton that will now be constructed
%accepts 
%does not recognize exactly the language $L_1$.
%to do so it would have to check the syntax of circuits,
%which weou

The automaton uses 11 states,
and its set of states is defined as
%$Q=\{?\} \cup \{0, 1\}\times\{{}^{0}, {}^{1}, {}^{\nwarrow}, {}^{\nearrow}, \text{\textvisiblespace}\}$.
%These states shall be denoted by
$Q=\{?$,
$0^{0}$, $0^{1}$, $\qvleft{0}$, $\qvright{0}$, $0$,
$1^{0}$, $1^{1}$, $\qvleft{1}$, $\qvright{1}$, $1\}$.
The initial function is defined by
$I(a)=\qvright{0}$ and $I(b)=\qvleft{0}$,
while the set of accepting states is $F=\{1\}$.

\begin{figure}[hbt]
	\centerline{\includegraphics[scale=1.1]{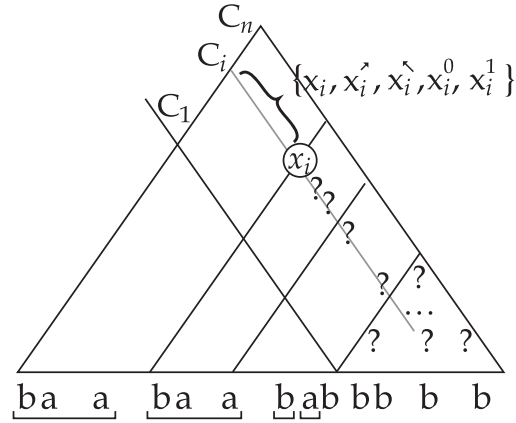}}
	\caption{Sketch of the computation.}
	\label{f:11_states_sketch}
\end{figure}

The overall structure of the computation of the automaton
on a valid encoding of a circuit
is given in Figure~\ref{f:11_states_sketch}.
The suffix $b^n$ of the encoding
is used by the automaton as the ``work space'',
and the diagonal spawned to the left
from every $i$th $b$ in this suffix
represents the computed value of the $i$th gate of the circuit.
Each diagonal initially holds the question mark;
in other words, $\Delta(wb^i)={?}$
for every sufficiently short suffix of the circuit description,
with the exception of $b$ and $\epsilon$.
The value of the $i$th gate is computed
on the substring starting at the description of the $i$th gate
and ending with $b^i$; formally,
\begin{equation*}
	\Delta(ba^{j_i} ba^{j_{i-1}} \ldots ba^{j_3} b ab b^i)
		=
	\left\{\begin{array}{rcl}
		\qv{0},
			& \text{if}
				& C_i=0; \\
		\qv{1},
			& \text{if}
				& C_i=1.
	\end{array}\right.
\end{equation*}
This computed value is propagated to the left,
so that all subsequent states in this diagonal are
$x^p \in Q$,
where $x \in \{0, 1\}$ is the value of the gate $C_i$,
while $p \in \{{}^{0}, {}^{1}, {}^{\nwarrow}, {}^{\nearrow}, \text{\textvisiblespace}\}$
is a state of an ongoing computation of the trellis automaton.

In order to compute the value of each $i$th gate,
the automaton should read the gate description $ba^{j_i}$
and look up the values of the gates $C_{j_i}$ and $C_{i-1}$,
which were computed on shorter substrings of the encoding
and are now being propagated in the diagonals.
To be more precise, the value of the gate $C_{j_i}$
should be brought to the $(i-1)$th diagonal
in the form of the state $\qvmem{x_{i-1}}{x_{j_i}}$,
and then the value of $C_i$
is computed and placed in the correct diagonal
by a single transition.

\begin{figure}[hbt]
	\centerline{\includegraphics[scale=1.1]{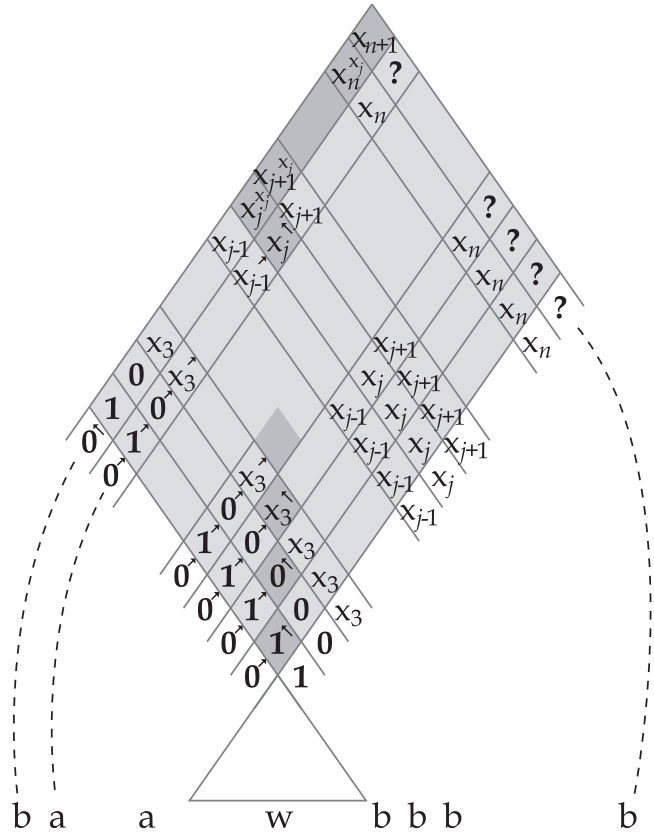}}
	\caption{Computing the value of the $i$th gate.}
	\label{f:11_states_gate_computation}
\end{figure}

The exact states of such a computation
are given in Figure~\ref{f:11_states_gate_computation}.
Assume that the encoding of the $(n+1)$th gate is $ba^j$
and it is propagated to the lower left border
of Figure~\ref{f:11_states_gate_computation}
in the form of the states $\qvright{0}$ for each $a$
and the state $\qvleft{0}$ for $b$.
The diagonals spawned from the $b^{n+1}$
arrive to the left as states
$\qv{x_i}$, $\qvmem{x_i}{0}$ or $\qvmem{x_i}{1}$
for each gate $i$,
and as $?$ for the last $(n+1)$-th gate.
Figure~\ref{f:11_states_gate_computation} illustrates
how the value of the $(n+1)$-th gate is computed,
while the already computed values of the rest of the gates are preserved.

\begin{figure}[hbt]
	\centerline{\includegraphics[scale=1.1]{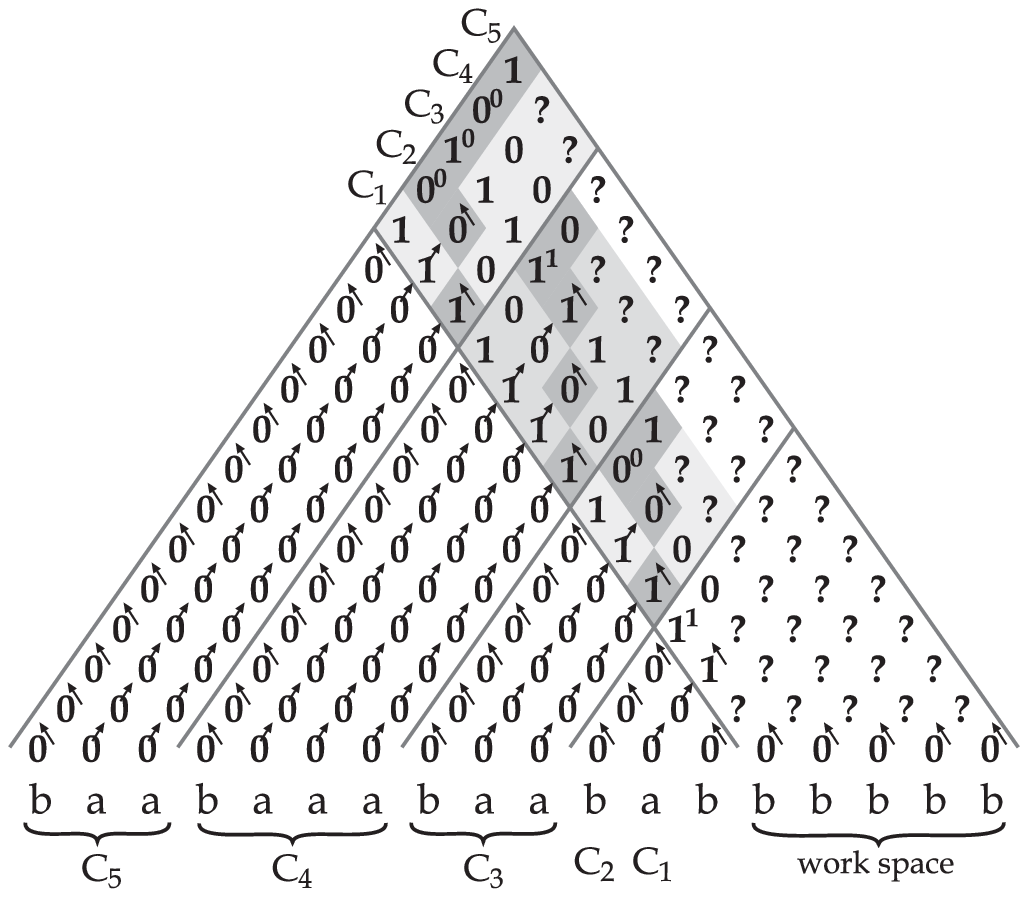}}
	\caption{A sample computation of the 11-state trellis automaton.}
	\label{f:11_states_computation}
\end{figure}

Furthermore, consider a full computation of the automaton
on a string $ba^2ba^3ba^2babb^5 \in L_1$,
given in Figure~\ref{f:11_states_computation}.
This computation contains three instances of computations
of the values of gates,
and each case is marked with dark grey
in the same way as in Figure~\ref{f:11_states_gate_computation}.

Now it is time to define all transitions
used in this computation.
The vertical line of states in $\{\qvleft{0}, \qvleft{1}\}$
marked with dark grey
represents matching the number of $a$s in the description of the gate
to the number of diagonals with gate values,
which allows seeking for the gate $C_j$.
This vertical line is maintained
by transitions of the form
\begin{equation*}
	\delta(\qvright{k}, \qv{\ell})=\qvleft{\ell}
		\quad (\text{for $k, \ell \in \{0, 1\}$}).
\end{equation*}
There are two cases of how this line can begin,
that is, how the bottom state $\qvleft{1}$ is computed.
If the previous gate $C_n$ refers to a gate other than $C_1$,
then the above general form of transitions\linebreak
gives $\delta(\qvright{0}, \qv{1})=\qvleft{1}$.
However, if $C_n$ is defined as $C_{n-1}=C_1$,
then the state $\qvmem{1}{1}$ will appear instead of $\qv{1}$
(this will be explained along with the below construction),
and the following extra transition is needed to handle this case:
\begin{equation*}
	\delta(\qvright{0}, \qvmem{1}{1})=\qvleft{1}.
\end{equation*}

The states to the left of this vertical line
belong to $\{\qvright{0}, \qvright{1}\}$,
and these states are computed by the following transitions:
\begin{equation*}
	\delta(\qvright{k}, \qvright{\ell})=\qvright{\ell}
		\quad (\text{for $k, \ell \in \{0, 1\}$}).
\end{equation*}
Beside the vertical line the transitions are:
\begin{equation*}
	\delta(\qvright{k}, \qvleft{\ell})=\qvright{\ell}
		\quad (\text{for $k, \ell \in \{0, 1\}$}).
\end{equation*}

Now consider the states
to the right of the dark grey vertical line,
which are all from $\{\qv{0}, \qv{1}\}$.
Beside the vertical line they are computed by the transitions
\begin{equation*}
	\delta(\qvleft{k}, \qv{\ell})=\qv{\ell}
		\quad (\text{for $k, \ell \in \{0, 1\}$}),
\end{equation*}
while further to the right the transitions are
\begin{equation*}
	\delta(\qv{k}, \qv{\ell})=\qv{\ell}
		\quad (\text{for $k, \ell \in \{0, 1\}$}).
\end{equation*}

All actual computations are done in the upper left border
of the area in Figure~\ref{f:11_states_gate_computation}.
Assume that the gate referenced by the gate $C_{n+1}$ is not $C_1$,
that is, $j \geqslant 2$ (as in the figure).
Then the transition in the leftmost corner of the area is
\begin{equation*}
	\delta(\qvleft{0}, \qvright{1})=\qv{1},
\end{equation*}
(note that this place is recognized by the automaton
because the value of $C_1$ is 1)
and the border continues to the up-right
by the transitions
\begin{equation*}
	\delta(\qv{k}, \qvright{\ell})=\qv{\ell}
		\quad (\text{for $k, \ell \in \{0, 1\}$}).
\end{equation*}
Eventually the upper left border
meets the dark grey vertical line,
which marks the diagonal corresponding to gate $C_j$.
The transition at this spot is
\begin{equation*}
	\delta(\qv{k}, \qvleft{\ell})=\qvmem{\ell}{\ell}
		\quad (\text{for $k, \ell \in \{0, 1\}$}),
\end{equation*}
and thus the value $\ell$ of the $j$-th gate is put to memory.
This memory cell is propagated in the up-right direction
by the transitions
\begin{equation*}
	\delta(\qvmem{k}{\ell}, \qv{m})=\qvmem{m}{\ell}
		\quad (\text{for $k, \ell, m \in \{0, 1\}$}).
\end{equation*}
This continues until the question mark in the $(n+1)$-th diagonal is encountered,
when the value of the $(n+1)$-th gate can be computed
by the following transition
\begin{equation*}
	\delta(\qvmem{k}{\ell}, ?)=\lnot(k \lor \ell) \in \{\qv{0}, \qv{1}\}
		\quad (\text{for $k, \ell \in \{0, 1\}$}).
\end{equation*}

Otherwise, if the $(n+1)$th gate refers to the gate $C_1$,
then the transition in the left corner of the figure is
\begin{equation*}
	\delta(\qvleft{0}, \qvleft{1})=\qvmem{1}{1},
\end{equation*}
which immediately concludes the dark grey vertical line.
The rest of the computation is the same as in the above description.

Having described the contents
of the upper left border of the area,
it is now easy to give the transitions
that compute its lower right border,
as these states are computed on the basis
of the upper left border of the computation for $C_n$.
If $C_n$ refers neither to $C_1$ nor to $C_2$,
then, as shown in the figure,
the second state in the lower right border
is computed by the transition
$\delta(\qvleft{1}, \qv{0})=\qv{0}$,
which has already been defined.
If $C_n$ refers to $C_1$,
then there will be a state $\qvmem{0}{1}$ instead of $\qv{0}$,
and if $C_n$ refers to $C_2$, there will be $\qvmem{0}{0}$ in this position,
so the following transitions are necessary:
\begin{equation*}
	\delta(\qvleft{1}, \qvmem{0}{k})=\qv{0}
		\quad (\text{for $k \in \{0, 1\}$}).
\end{equation*}
The rest of the states in the lower right border
are either computed by the earlier defined transitions
$\delta(\qv{k}, \qv{\ell})=\qv{\ell}$,
or by the transitions
\begin{equation*}
	\delta(\qv{k}, \qvmem{\ell}{m})=\qv{\ell}
		\quad (\text{for $k,\ell,m \in \{0, 1\}$}).
\end{equation*}
This completes the list of transitions
used to compute the value of each gate starting from $C_3$.
A few more transitions are required to initialize the computation
and to set the values of $C_1$ and $C_2$.

Each symbol $b$ in a gate description $ba^j$
is propagated in the right-up direction
by the transition
\begin{equation*}
	\delta(\qvleft{0}, \qvright{0})=\qvleft{0}.
\end{equation*}
The question marks are created from any two subsequent $b$s
by the transition
\begin{equation*}
	\delta(\qvleft{0}, \qvleft{0})={?}.
\end{equation*}
The question marks are reduplicated by the transitions
\begin{equation*}
	\delta(q, ?)={?}
		\quad (\text{for $q \in \{?, \qv{0}, \qv{1}\}$}),
\end{equation*}
and by one more transition
that works in the case of $C_{n+1}=C_n \downarrow C_n$:
\begin{equation*}
	\delta(\qvright{0}, ?)={?}.
\end{equation*}

\begin{figure}[hbt]
	\centerline{\includegraphics[scale=1.1]{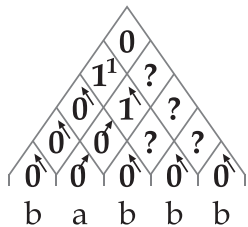}}
	\caption{The beginning of the computation.}
	\label{f:11_states_computation_begins}
\end{figure}

The beginning of the computation
is illustrated in Figure~\ref{f:11_states_computation_begins}:
as every valid circuit description
has a substring $babbb$,
these transitions are needed in every computation.
Here the value of $C_1$ is set by the transition
\begin{equation*}
	\delta(\qvright{0}, ?)=\qvleft{1},
\end{equation*}
while processing the gate $C_2$
requires the transition
\begin{equation*}
	\delta(\qvleft{1}, ?)={?}.
\end{equation*}
This concludes the description of the transition function.
To make it total,
the rest of the transitions can be defined arbitrarily.

Some transitions defined above will actually never occur.
Note that no sequential NOR circuit
may have two consecutive gates with value 1:
if $C_n=1$, then $C_{n+1}=\lnot(C_n \lor C_{j_{n+1}})=\lnot 1=0$.
This makes the transitions
$\delta(q, q')$ with
$q, q' \in \{\qv{1}, \qvright{1}, \qvleft{1}, \qvmem{1}{0}, \qvmem{1}{1}\}$
impossible,
and as 11 such transitions have been defined above,
they may be safely undefined (or redefined arbitrarily).
With this correction,
the transition table of the automaton
is given in Table~\ref{t:11_states_delta}.

\begin{table}[hbt]
	\centerline{$\begin{array}{|c|ccccccccccc|}
\hline
		& \qunknown	& \qv{0}	& \qv{1}	& \qvright{0}	& \qvright{1}	& \qvleft{0}	& \qvleft{1}	& \qvmem{0}{0}	& \qvmem{0}{1}	& \qvmem{1}{0}	& \qvmem{1}{1} \\
\hline
\qunknown	& \qunknown	& 	& 	& 	& 	& 	& 	& 	& 	& 	&  \\
\qv{0}	& \qunknown	& \qv{0}	& \qv{1}	& \qv{0}	& \qv{1}	& \qvmem{0}{0}	& \qvmem{1}{1}	& \qv{0}	& \qv{0}	& \qv{1}	& \qv{1} \\
\qv{1}	& \qunknown	& \qv{0}	& 	& \qv{0}	& 	& \qvmem{0}{0}	& 	& \qv{0}	& \qv{0}	& 	&  \\
\qvright{0}	& \qvleft{1}	& \qvleft{0}	& \qvleft{1}	& \qvright{0}	& \qvright{1}	& \qvright{0}	& \qvright{1}	& 	& 	& 	& \qvleft{1} \\
\qvright{1}	& 	& \qvleft{0}	& 	& \qvright{0}	& 	& \qvright{0}	& 	& 	& 	& 	&  \\
\qvleft{0}	& \qunknown	& \qv{0}	& \qv{1}	& \qvleft{0}	& \qv{1}	& \qunknown	& \qvmem{1}{1}	& 	& 	& 	&  \\
\qvleft{1}	& \qunknown	& \qv{0}	& 	& 	& 	& 	& 	& \qv{0}	& \qv{0}	& 	&  \\
\qvmem{0}{0}	& \qv{1}	& \qvmem{0}{0}	& \qvmem{1}{0}	& 	& 	& 	& 	& 	& 	& 	&  \\
\qvmem{0}{1}	& \qv{0}	& \qvmem{0}{1}	& \qvmem{1}{1}	& 	& 	& 	& 	& 	& 	& 	&  \\
\qvmem{1}{0}	& \qv{0}	& \qvmem{0}{0}	& 	& 	& 	& 	& 	& 	& 	& 	&  \\
\qvmem{1}{1}	& \qv{0}	& \qvmem{0}{1}	& 	& 	& 	& 	& 	& 	& 	& 	&  \\
\hline
\end{array}
$}
	\caption{The transition table of the 11-state trellis automaton.}
	\label{t:11_states_delta}
\end{table}

The correctness of the given construction
is stated in the following lemma,
which specifies the state
computed on (almost) every substring
of a valid encoding of a circuit.

\begin{lemma}\label{eleven_states_TA_correctness_lemma}
Let $w b^n$ with $w \in \{a, b\}^*$ and $n \geqslant 2$
be a description of a circuit
with the values of gates $x_1, \ldots, x_n \in \{0, 1\}$.
Then:
\begin{enumerate}\renewcommand{\theenumi}{\roman{enumi}}
\item	\label{eleven_states_TA_correctness_lemma__claim_wci}
	$\Delta(w b^i) \in \{x_i, \qvmem{x_i}{0}, \qvmem{x_i}{1}\}$
	for $1 \leqslant i \leqslant n$, and $\Delta(w b^n)=x_n$.
\item	\label{eleven_states_TA_correctness_lemma__claim_uwcn}
	$\Delta(u w b^n) \in \{x_n, x_n^0, x_n^1, \qvleft{x_n}, \qvright{x_n}\}$
	for every $u \in \{a, b\}^*$;
\item	\label{eleven_states_TA_correctness_lemma__claim_aiwcj}
	$\Delta(a^i w b^j) = \left\{\begin{array}{ccl}
		\qvright{x_i} & \text{if} & j<i, \\
		\qvleft{x_i} & \text{if} & j=i, \\
		x_i & \text{if} & j>i.
	\end{array}\right.
	\quad (1 \leqslant i < n, \; 1 \leqslant j \leqslant n)$;
\item	\label{eleven_states_TA_correctness_lemma__claim_baiwcj}
	$\Delta(ba^i w b^j) = \left\{\begin{array}{ccl}
		x_i & \text{if} & j<i, \\
		\qvmem{x_j}{x_i} & \text{if} & j \geqslant i.
	\end{array}\right.
	\quad (1 \leqslant i < n, \; 1 \leqslant j \leqslant n)$
\end{enumerate}
\end{lemma}

A formal proof is omitted,
as every transition has been explained along with the construction.
It could be carried out by an induction on the length of $w$.

This establishes the main result of this paper:

\begin{theorem}\label{eleven_states_theorem}
There exists an 11-state trellis automaton
with 50 useful transitions
that recognizes a \family{P}-complete language
over a 2-letter alphabet.
\end{theorem}

This automaton can be converted
to a linear conjunctive grammar,
which has a nonterminal representing every state
and at most 4 rules for each transition.

\begin{corollary}
There exists a linear conjunctive grammar
with 11 nonterminals
and at most 200 rules
that recognizes a \family{P}-complete language
over a 2-letter alphabet.
\end{corollary}

Although this grammar is significantly smaller
than the earlier example~\cite{LinearHardest},
it is still large.
However, it is conjectured that
the principles of the operation of this trellis automaton
can be implemented in a linear conjunctive grammar
much more efficiently,
and a much smaller grammar
generating (almost) the same language
can be obtained.

\section{Further work}

The result on the existence
of an 11-state trellis automaton
recognizing a \family{P}-complete language
could (in theory) be improved in several ways.

One possibility is that some encoding
of the same Sequential NOR Circuit Value Problem,
perhaps the very same encoding,
could be recognized by an automaton with 10 states of fewer.
Constructing such an automaton
would be a challenging exercise in programming,
though it would not give any new knowledge
on \family{P}-completeness as such.

Perhaps a more promising direction is to try to invent
a different \family{P}-complete problem and its encoding,
which would admit a solution
by a significantly smaller trellis automaton.
Such a problem would be interesting in itself,
% that we shall learn something new about ta
and in this way a search for a small automaton
would become more than just an exercise.

\begin{comment}
Finally, one should still bear in mind
that all these results are 
If it turns out that DLOGSPACE=P
and all the theory of P-completeness is trivial,
then, in particular, the smallest trellis automaton
solving a P-complete problem
would contain 1 state.
\end{comment}

\bibliographystyle{eptcs}

\end{document}